\documentclass[sigconf]{acmart}

\usepackage{xspace}
\usepackage{multirow}
\usepackage{enumitem}
\usepackage{stfloats}
\usepackage{subcaption}
\usepackage{makecell}
\usepackage{soul}
\usepackage{tipa}

\AtBeginDocument{%
  }

\renewcommand\footnotetextcopyrightpermission[1]{} 

\copyrightyear{2025}
\acmYear{2025}
\setcopyright{acmlicensed}\acmConference[UbiComp Companion '25]{Companion of the 2025 ACM International Joint Conference on Pervasive and Ubiquitous Computing}{October 12--16, 2025}{Espoo, Finland}
\acmBooktitle{Companion of the 2025 ACM International Joint Conference on Pervasive and Ubiquitous Computing (UbiComp Companion '25), October 12--16, 2025, Espoo, Finland}
\acmDOI{10.1145/3714394.3754429}
\acmISBN{979-8-4007-1477-1/2025/10}

\definecolor{lightblue}{RGB}{61,201,234} 
\definecolor{darkblue}{RGB}{49,160,204}

\newcommand{\trackrevnice}[1]{#1}

\newcommand{\todocite}[1]{\textcolor{red}{TODO: cite}}
\newcommand{\redacted}[1]{\textbf{REDACTED}}
\newcommand*{\projectname}{HEar-ID\xspace}

\begin{document}\sloppy

\title[Turning Your Ear into Biometric Key]{Poster: Recognizing Hidden-in-the-Ear Private Key for Reliable Silent Speech Interface Using Multi-Task Learning}

\author{Xuefu Dong}
\author{Liqiang Xu}
\affiliation{
  \institution{The University of Tokyo}
  \state{Tokyo}
  \country{Japan}
  }


\author{Lixing He}
\affiliation{
  \institution{The Chinese University of Hong Kong}
  \state{Hong Kong SAR}
  \country{China}
  }
\email{}

\author{Zengyi Han}
\authornote{Zengyi Han is the contact author.\\
Posted for personal use only; not for redistribution. The definitive version will appear in UbiComp Companion ’25. https://doi.org/10.1145/3714394.3754429}
\affiliation{
  \institution{Dalian Maritime University}
  \city{Dalian}
  \state{}
  \country{China}
  }
\email{zyhan@dlmu.edu.cn}

\author{Ken Christofferson}
\affiliation{%
  \institution{University of Toronto}
  \city{Toronto}
  \state{Ontario}
  \country{Canada}}
  
\author{Yifei Chen}
\affiliation{
  \institution{Tsinghua University}
  \state{Beijing}
  \country{China}
}

\author{Akihito Taya}
\affiliation{
  \institution{The University of Tokyo}
  \state{Tokyo}
  \country{Japan}
  }

\author{Yuuki Nishiyama}
\affiliation{
  \institution{The University of Tokyo}
  \state{Chiba}
  \country{Japan}
  }

\author{Kaoru Sezaki}
\affiliation{
  \institution{The University of Tokyo}
  \state{Chiba}
  \country{Japan}
  }

\renewcommand{\shortauthors}{Xuefu Dong et al.}

\begin{abstract}

Silent speech interface (SSI) enables hands‐free input without audible vocalization, but most SSI systems do not verify speaker identity. We present HEar-ID, which uses consumer active noise–canceling earbuds to capture low‐frequency “whisper” audio and high‐frequency ultrasonic reflections. Features from both streams pass through a shared encoder, producing embeddings that feed a contrastive branch for user authentication and an SSI head for silent spelling recognition. This design supports decoding of 50 words while reliably rejecting impostors, all on commodity earbuds with a single model. Experiments demonstrate that HEar-ID achieves strong spelling accuracy and robust authentication.

\end{abstract}



\begin{teaserfigure}
\centering
\includegraphics[width=0.9\linewidth]{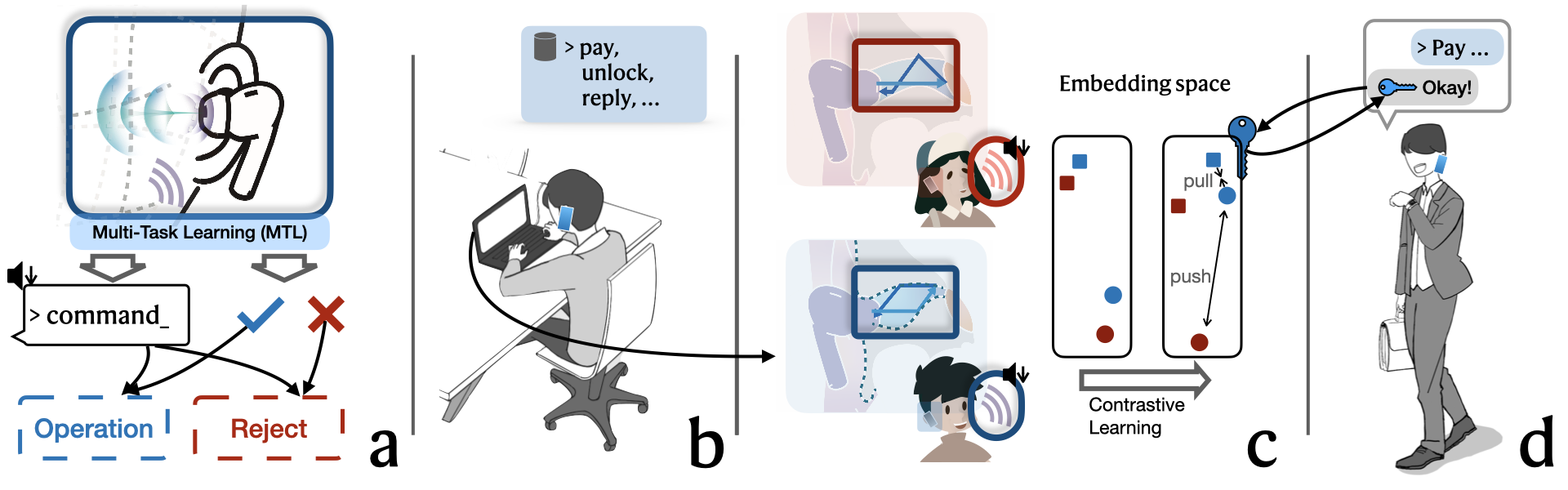}
\vspace{-1em}
\caption{(a) \projectname uses multi-task learning (MTL) to authenticate user identity and reliably infer silent speech; (b) a new user registers in \projectname by recording few shots of silent commands; (c) \projectname leverages contrastive learning to pull cosine similarity between projected features of whisper and ECDM; (d) the user can be verified in a natural and safe style.}
\label{fig:teaser}
\end{teaserfigure}

\maketitle

\section{Introduction} \label{sec:introduction}

Although speech‐based interactions offer a natural alternative to manual or visual controls, they are not always feasible in public settings: 
speaking aloud in quiet environments (such as a library) may disturb others or expose private information (e.g., dictating a text message). 
Silent speech interface (SSI) addresses these issues by relying on mouth and facial movements rather than audible vocalization, allowing use in both noisy and quiet spaces without disrupting bystanders. Consequently, users tend to prefer SSI over traditional speech recognition in public contexts~\cite{pandey2021acceptability}. 
However, prior SSI-related works ~\cite{earssr, zhang2023echospeech, magnetometer_ssr, juntao2025m2silent} hardly consider the safety issues. 
While normal voice‐based authentication mitigates these risks by verifying the speaker’s identity before granting access, 
it can be vulnerable to replay and injection attacks (e.g., triggering Siri via a loudspeaker),
leaving an imperative need for developing a reliable SSI system.



We analyzed and found that the silent speech recognition task and the speaker authentication task are correlated rather than independent. The inherent structure uniqueness of each individual's ear canal creates distinct acoustic propagation paths, so that subtle ear canal deformations encode both the utterance content and speaker identity.
\trackrevnice{Fortunately, recent years have witnessed an essential boost in earable sensing ranging from activity recognition~\cite{he2025embodiedsense, lyu2024earda, shimojo23earble}, pose or mesh reconstruction~\cite{li2022eario, duan2025argus, imuposer, gao2025expression}, speech enhancement~\cite{he2023towardsvibvoice, headsetmic_hiraki, whispermask_hiraki}, and user authentication~\cite{duan2024f2key, wang2021eardynamic, gao2019earecho, fu_100p_imuCardiac}. Among them, several works use commodity earbuds with an in-ear microphone to detect ear canal dynamic motion (ECDM) using ultrasonic sensing, where the ear canal serves as the reflection chamber. 
}
Therefore, we employ a multi‐task learning (MTL) approach upon the earable platform, leveraging this correlation to perform silent spelling recognition and speaker authentication simultaneously.

In this work, we enable reliable SSI 
by proposing \projectname, which only leverages a commodity active noise‐canceling earbud to emit an inaudible OFDM signal and record both ultrasonic reflections and whisper audio to enable silent spelling input (e.g. \textipa{/i: eI Ar/}) for the word "ear") and user verification with a single machine learning model. 
Here we treat the silent speech as equivalent to whispering, since we found most experiment participants still make a subtle voice due to a lack of training.  
\trackrevnice{As shown in Figure~\ref{fig:teaser}, \projectname extracts features from ultrasonic waveforms (17.5–23kHz) and whispers (0-11kHz), then feeds the twin features through a shared encoder (TCN→Bi‐GRU→MLP) to produce $d$‐dimensional embeddings. }
A contrastive head aligns genuine‐user whisper–ultrasonic pairs and repels attacker samples, while an SSI head on concatenated embeddings enables word-level decoding of silent spellings. In the preliminary experiments, \projectname consistently delivered promising results: for 11 participants, the system reliably rejected impostors with a false positive rate (FPR) of 3.2\% and a true positive rate (TPR), accompanying 90.25\% Top‐1 word recognition accuracy for eight of them.

Our contributions are: (1) a multi‐task SSI framework that fuses ultrasonic features with whisper ones for joint spelling and authentication; (2) a contrastive learning formulation (CLWUM) that embeds whisper and ultrasonic signals into a “private‐key” space for robust verification; (3) an evaluation demonstrating accuracy spelling on 50 words and strong authentication.

\section{Related Work}
\subsection{Ultrasonic Silent Speech Interfaces}
Ultrasonic sensing interpreted by doppler effect~\cite{chuanxin_prl} or other machine learning techniques ~\cite{mao2020deeprange, cheng2023twinkletwinkle, cheng2022pdfmcw} detects movements similar to those captured by radio-frequency-~\cite{wenwei2025rethinking, zeng2023msilent}, optics-~\cite{su2023liplearner, wei2024uniqr} and motion-based sensors~\cite{han2024rideguard,he2025vibomni} to enable silent speech interface (SSI) techniques.
Wearable approaches include 
\citet{jin2022earcommand}’s EarCommand which uses a custom earbud emitting an FMCW chirp, with a microphone in the ear canal recording reflections, achieving 89.9\% accuracy on 32 silently spoken words across 12 participants.

\subsection{Earable‐based User Authentication}
Wearable authentication can use passwords—e.g., voice or gesture‐based schemes on earbuds~\cite{chen2024enabling,Xu2020EarBuddyEO}—but spoken passwords risk leaking personal data and are vulnerable to replay attacks, motivating biometric alternatives. 

However, almost all existing systems require normal speech or detectable motion, limiting usability in public or quiet settings. To our knowledge, \projectname is the first earable solution to authenticate via silent or whispered speech. Unlike EarDynamic \cite{wang2021eardynamic}, which relies solely on ultrasonic reflections, we learn a joint mapping between ultrasonic and whisper signals to capture user identity even when no audible speech is present.



\section{Contrastive Multi-Task Learning (CMTL)}
\begin{figure*}
    \centering
    \includegraphics[width = 0.75\textwidth]{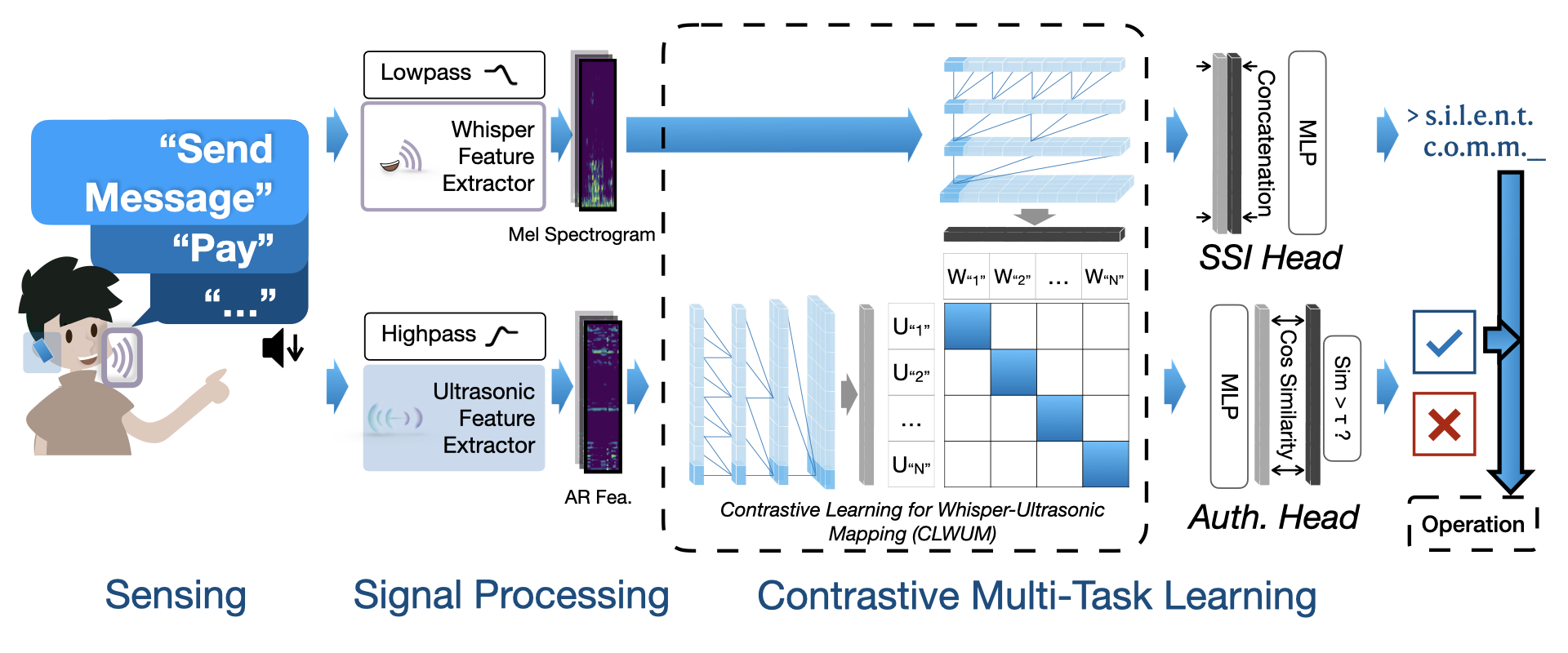}
    \vspace{-1.5em}
    \caption{The three-fold workflow of \projectname.}
\vspace{-1em}
    \label{fig:workflow}
\end{figure*}

In this section, we describe \projectname's approach for word inference and user verification,
as illustrated in \autoref{fig:workflow}.

\subsection{Signal Processing and Feature Extraction}
\label{sec:feature_extraction}
\projectname leverages an OFDM signal of 2046 samples over a sampling rate of 48000Hz, lasting for about 42ms. 
Upon receiving the reflection from the ear canal, \projectname 
first isolates the \textit{ultrasonic band} (17.5--23\,kHz) with a high‐pass filter, then aligns each received segment to an exact OFDM cycle. A coarse alignment finds the lag \(t_{\mathrm{coarse}}\) that maximizes cross‐correlation between one transmitted symbol and two received symbols. A fine alignment then corrects sub‐sample delays (±25 samples) by minimizing phase differences with the transmitted signal. The original audio is aligned using the result index. 

We extract features from the data after slicing it into 426-ms (10 frames) sliding windows with a stride of 85ms (2 frames). The \textit{ultrasonic band} of each window, containing information of ear canal dynamic motions (ECDM), is further boosted by a second‐order differentiation for compensation of rapid attenuation above 16\,kHz in commercial earbuds, and later represented by 200-lag autoregressive (AR) coefficients.
On the other hand, the whispering audio is procured by applying another 12kHz lowpass filter to the \textit{aligned original audio}. To reduce computational complexity, we downsample the audio by a factor of 2. 
Then we fed it into a widely adopted mel-spectrogram extractor ($n\_fft = hop\_length \approx 42ms$). The paired features are then fed into our neural network.


\subsection{Contrastive Learning for Whisper-Ultrasonic Mapping (CLWUM)}

When a user articulates silently, facial movements induce correlated perturbations in both the low‐frequency whisper signal (captured by an earbud’s microphone) and the high‐frequency ultrasonic reflections inside the ear canal. CLWUM exploits this intrinsic link by mapping paired whisper and ultrasonic signals into a shared embedding space, effectively acting as a “private key” that aligns only genuine‐user embeddings.

As illustrated in Figure~\ref{fig:workflow}, each batch provides $2N$ segments per modality: the first $N$ from the \emph{genuine} user and the next $N$ from \emph{attackers} (N is set to 50 words in experiments). After modality‐specific encoders $f_w(\cdot)$ (whisper) and $f_u(\cdot)$ (ultrasonic) and small projection heads $h_w(\cdot)$, $h_u(\cdot)$, we obtain embeddings
\[
w_i = h_w\bigl(f_w(x_i^{\mathrm{whisper}})\bigr),\quad
u_j = h_u\bigl(f_u(x_j^{\mathrm{ultra}})\bigr),
\quad i,j=1,\dots,2N,
\]
which we collect into $W(d\times 2N)$, and $U(d\times 2N)$
Since CLWUM focuses on the genuine‐user mapping, we extract only the first $N$ embeddings from the genuine user:
$
W^g (d\times N), 
\quad
U^g (d\times N).
$
We then compute the cosine‐similarity array between whisper and ultrasonic embeddings from $N$ genuine user samples $S_{ij}=sim(w_i,\,u_j)$
Here, the diagonal entries $S_{ii}$ represent true whisper–ultrasonic pairs.  Contrastive learning fosters a robust cross‐modal alignment between ultrasonic and whisper embeddings for silent spelling recognition, while simultaneously extracting the shared features that underpin reliable user authentication.
Accordingly, we define the contrastive loss:
\[
L_{\mathrm{CL}}
= -\frac{1}{N}\sum_{i=1}^N
\log 
\frac{\exp\!(S_{i,i}/\tau)}
     {\sum_{j=1}^{N} \exp\!(S_{i,j}/\tau)},
\]
where $\tau>0$ is a temperature hyperparameter and is set to 0.7 based on empirical experience. Minimizing $L_{\mathrm{CL}}$ maximizes similarity of genuine pairs and minimizes similarity of mismatched pairs, embedding whisper and ultrasonic signals from the genuine user into a coherent shared space.

\subsection{MTL-based Reliable Silent Spelling Interface}
To simultaneously verify user identity and recognize spelled words in silent speech, we adopt a multi-task learning (MTL) paradigm. Multi-task learning aims to enhance the overall performance on
multiple tasks by leveraging shared representation and useful information across related tasks \cite{zhang2018overview}. Our architecture branches into two tasks—User Authentication and Silent Spelling Recognition—sharing the CLWUM encoders as a foundation.

\subsubsection{User Authentication}

As shown in Figure~\ref{fig:workflow}, each whisper or ultrasonic segment first passes through the CLWUM encoder—three stacked TCN blocks followed by two Bi‐GRU layers—and then through an embedding MLP to produce a $d$-dimensional representation. This shared embedding is further projected by lightweight modality‐specific heads to yield authentication vectors. For each batch, let $h_i^{\mathrm{whisper}}$ and $h_i^{\mathrm{ultra}}$ be the shared $d$-dimensional embeddings (after CLWUM encoder and embedding MLP) corresponding to the $i$-th genuine‐user silent spelling, and let $h_j^{\mathrm{ultra}}$ be the embedding for the $j$-th attacker segment (where $i,j=1,\dots,N$ and there are $N$ unique words per batch). We then compute the authentication vectors by applying two separate MLP for ultrasonic and whisper data in the authentication head. The system then calculates the cosine similarity of the pair, and makes the dicision of authentication if similarity is greater than a threshold $thr$. During training, we form pairs in dataset from genuine-user and attackers:
\[
w_i = \mathrm{MLP}_w\bigl(h_i^{\mathrm{whisper}}\bigr),\quad
u_i^+ = \mathrm{MLP}_u\bigl(h_i^{\mathrm{ultra}}\bigr),\quad
u_j^- = \mathrm{MLP}_u\bigl(h_j^{\mathrm{ultra}}\bigr),
\]
where $w_i,u_i^+,u_j^-\in\mathbb{R}^d$. Here, $(w_i, u_i^+)$ form genuine‐user pairs (both embedding the same word), while each $u_j^-$ represents an attacker embedding for that same word.

Then, we compute cosine similarities 
$s_i^+ =sim(w_i,\,u_i)$ measures similarity for a genuine pair, and $\ s_{i}^- =sim(w_i,\,u_{i^*}^-)$ for a genuine–attacker pair. 
Training minimizes the angular triplet loss with margin m (where m is set to 11.45):
\[
L_{\mathrm{auth}}
= \frac{1}{N}\sum_{i=1}^N
\max\Bigl(0,\;m + arccos(s_i^+) - arccos(s_{i}^-)\Bigr).
\]
This loss pulls the angle between genuine pairs smaller (higher similarity) while pushing the hardest negative (attacker) angle larger, yielding a clear decision boundary. We calibrate the $thr$ with Youden's J statistics~\cite{youden1969statistical}, and adopt the empirical value of m as 30 degrees.

\subsubsection{Silent Spelling Recognition}

As depicted in Figure~\ref{fig:workflow}, for silent spelling the genuine‐user whisper and ultrasonic segments are first processed by the shared CLWUM encoder (TCN $\to$ Bi‐GRU $\to$ embedding MLP), resulting in two sequences of $d$-dimensional vectors:
$
h^{\mathrm{whisper}} = \bigl[h_1^w, \dots, h_N^w\bigr]$,$
h^{\mathrm{ultra}} = \bigl[h_1^u, \dots, h_N^u\bigr].
$
We then concatenate corresponding whisper and ultrasonic embeddings for each time step to $c_i = \bigl[h_i^w \,\Vert\, h_i^u\bigr]$
and feed the concatenated sequence $\{c_i\}$ into a small projection MLP followed by a softmax classifier to produce frame-level logits over the alphabet (26 letters + blank):
$
\ell_i = \mathrm{MLP}_{\mathrm{spell}}(c_i),
$ 
where $K=27$. Denote the entire logit sequence by $\mathbf{L} = [\ell_1, \dots, \ell_N]$.

We train with Connectionist Temporal Classification (CTC) loss~\cite{graves2012connectionist} to align $\mathbf{L}$ with the target letter sequence $\mathbf{y}$ without frame‐wise labels:
$L_{\mathrm{CTC}} = -\log P(\mathbf{y} \mid \mathbf{L}).$
CTC allows the model to learn both letter identities and their temporal transitions implicitly, which is crucial for decoding silently mouthed words.

\subsection{Authentication Pipeline}
\label{sec:training_inference}

The authentication pipeline is divided into two phases:

\noindent
\textbf{Registration Phase.} 
Each user begins by silently spelling a small set of word samples while wearing the earbuds. \projectname then trains the network using whisper–ultrasonic pair of samples with other individuals' data. We claim that these extra training data can be provided by the manufacturer along with the product. 
\projectname is trained end-to-end by minimizing the combined loss
\[
L_{\mathrm{total}} = \alpha\,L_{\mathrm{CL}} + \beta\,L_{\mathrm{auth}} + \gamma\,L_{\mathrm{CTC}},
\]
with task weights set to $\alpha=0.1$, $\beta=0.5$, and $\gamma=0.3$ after hyperparameter tuning.
A verification threshold is then calibrated. 

\noindent
\textbf{Verification Phase.}
When a user silently spells a word, the system captures the corresponding whisper and ultrasonic segments, computes their embeddings, and compares these against the stored reference set. If the similarity to at least one reference embedding exceeds the threshold, the user is accepted; otherwise, access is denied. At the same time, the concatenated embeddings are sent to the head, which produces logits that are decoded (e.g., via beam search) to yield the final spelled word. 

\section{Preliminary Results}

\begin{figure*}[tb!]
\centering
\hbox{\subfloat{\label{fig: word}
\includegraphics[height=0.15\textwidth]{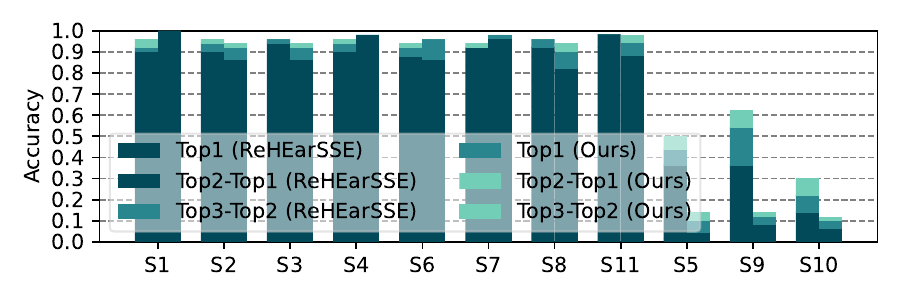}}
\subfloat{\label{fig: auth}
\includegraphics[height=0.14\textwidth]{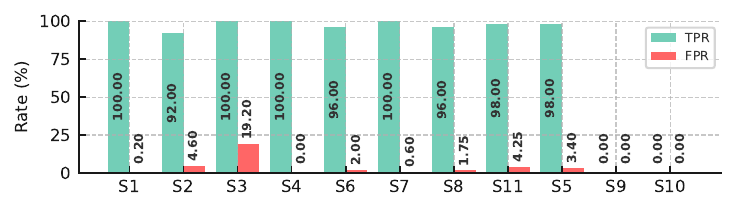}}}
\vspace{-1em}
\caption{(a) Word inference accuracy, and (b) user authentication performance.}
\vspace{-1em}
    \label{fig: overall}
\end{figure*}

We used a subset of the dataset from ReHEarSSE~\cite{dong2024rehearsse} 
(10 participants) and recruited one additional participant, bringing our study to a total of 11 users (9 males, 2 females; mean age 22).
All data is collected by off-the-shelf active noise-cancelling earbuds (Edifier W380NB) where we removed the in-ear mic from the earbud circuitry and wired it directly to a smartphone (Google Pixel 3a) via a 3.5 mm jack, ensuring unprocessed raw audio.
All data collection procedures were conducted in accordance with Institutional Review Board (IRB) guidelines.
All participants spelled 4 rounds of the same lexicon of 50 randomly selected words from the 1000 most used words from the Oxford English Dictionary ~\cite{oxford2023EngDic} with subtle voice (e.g. \textipa{/i: eI Ar/}) for the word "ear").

\subsection{Overall Performance}
\subsubsection{Experiment Design:}
We take turns to regard one participant as the genuine user and others as impostors. For each participant, we form the testing set by randomly choosing a re-wearing session, and the training set by all other sessions except the left-out session. 
The authentication head and the CLWUM are trained and tested on all participants' data. That means each participant's model is attacked 500 times and accessed by the genuine user 50 times during testing.
For the SSI head, we only use the genuine user's data, resulting in a user-dependent model. 
We implement the pipeline of ~\cite{dong2024rehearsse} serving as a baseline.

\subsubsection{Metrics.}
For recognition, we report \emph{Top–$n$ word accuracy} ($n\!=\!1,2,3$), i.e.\ the percentage where the true label appears in the top-$n$ predictions.
For authentication, we adopt the standard \textbf{TPR} (True Positive Rate), \textbf{FPR} (False Positive Rate) and the resulting \textbf{EER}.  Figure\,\ref{fig: overall} illustrates the recognition accuracy of our \textbf{whisper+ultrasonic} Silent-Speech Interface (SSI) with the ultrasonic-only baseline ReHEarSSE~\cite{dong2024rehearsse}, while Figure\,\ref{fig: overall} summarises per–user authentication results (TPR/FPR).%

\noindent\textbf{Silent-speech recognition:}  
Our \emph{whisper\,+\,ultrasonic} SSI attains a mean Top-1 accuracy of \textbf{67.3\,\%}. However, we found that for 8 of 11 users without degraded performance in ultrasonic sensing, we achieved 90.25\% Top-1 accuracy, which is comparable to the ultrasonic-only \textsc{ReHEarSSE} baseline (\textbf{91.4\,\%}). Three speakers (S1, S4, S7) exceed 90 \%, confirming the benefit of the whisper stream. S5, S9, and S10 lag (<10 \%)—likely owing to unclear articulation and idiosyncratic sensor placement, issues that the whisper–ultrasonic alignment amplifies.

\noindent\textbf{Authentication:}  
Average performance is \textbf{TPR = 81.76\,\%}, \textbf{FPR = 3.2\,\%}. Notably, S5 achieves \textbf{99.9\,\% / 3.4\,\%} despite poor recognition, illustrating that our identity embeddings remain robust when lexical decoding falters. Conversely, S9 and S10 record near-zero TPRs, mirroring their low recognition scores and indicating inconsistent headset placement across sessions.  


Overall, the results confirm that coupling a whisper microphone with ultrasonic sensing brings recognition accuracy on par with state-of-the-art acoustic SSI while preserving high authentication robustness—meeting the dual goals of content retrieval and user verification in a single wearable and machine learning model.

\section{Conclusion}
\balance 

We introduced \projectname, which combines AR ultrasonic and whisper features to enable silent spelling and user authentication on earbuds. A CLWUM module aligns genuine‐user embeddings and repels impostors to simultaneously support two tasks. Tests show robust authentication under real‐world conditions and promising spelling accuracy of 90.25\% over 8 out of 11 participants for 50 words. Future work will expand lexicons and refine continuous verification, leveraging methods including a generative model to produce realistic synthetic data~\cite{gong2025data, fu2022svoice}.



\begin{acks}
This work was supported by JSPS KAKENHI Grant Number JP24K20759,
 by the Fundamental Research Funds for the Central Universities (\href{tel:3132025240}{3132025240}), China.
\end{acks}

\bibliographystyle{ACM-Reference-Format}
\bibliography{refs}

\end{document}